\documentclass{aa}
\usepackage{natbib}
\bibpunct{(}{)}{;}{a}{}{,}

\usepackage{epsfig}
\usepackage{graphics}
\usepackage{float}
\usepackage{amsmath}
\usepackage{multirow}
\usepackage{longtable}
\usepackage{rotate}
\usepackage{subfigure}
\def\sidehead#1{\noalign{\vskip 1.5ex}\multicolumn{4}{@{}l}{\em #1}\\
                \noalign{\vskip .5ex}}

\def\phn{\phantom{0}}  
\def\phs{\phantom{$-$}}    

\def\tablecomments#1{\par\smallskip\noindent Notes. #1}
\def\plotone#1{\centerline{\psfig{figure=#1,width=\hsize,clip=}}}
\def\kms{\ifmmode{\rm km\,s^{-1}}\else\hbox{$\rm km\,s^{-1}$}\fi}
\def\nodata{\phs$\cdots$}

\newcommand{\Teff}{$\mathrm{T}_{\mathrm{eff}}$}

\let\arcdeg\degr
\let\simgt\ga

\setlongtables

\begin{document}

   \title{Pulsating or not? A search for hidden pulsations below the red edge of the ZZ 
    Ceti instability strip\thanks{The data presented herein were obtained
       at the W.M. Keck Observatory, which is operated as a scientific
       partnership among the California Institute of Technology, the
       University of California and the National Aeronautics and Space
       Administration. The Observatory was made possible by the generous
       financial support of the W.M. Keck Foundation.}}

   \author{R. Kotak\inst{1}
          \and{M. H. van Kerkwijk}\inst{2}
           \and{J. C. Clemens}\inst{3}\thanks{Alfred P. Sloan Research Fellow}
            \and{T. A. Bida}\inst{4}}

   \offprints{R. Kotak}

   \institute{Lund Observatory, Box 43, SE-22100 Lund, Sweden\\
              \email{rubina@astro.lu.se}
            \and
              Astronomical Institute, Utrecht University,
              P. O. Box 80000, 3508~TA Utrecht, The Netherlands\\
              \email{M.H.vanKerkwijk@astro.uu.nl}
             \and
              Department of Physics and Astronomy, University of
              North Carolina, Chapel Hill, NC 27599-3255, USA\\
              \email{clemens@physics.unc.edu}
            \and
              Lowell Observatory, 1400 W Mars Hill Rd.,
              Flagstaff, AZ 86001, U.S.A. \\
             \email{tbida@lowell.edu}
             }

   \date{Received ; accepted }

   \abstract{The location of the red edge of the ZZ Ceti instability strip is defined
             observationally as being the lowest temperature for which a white dwarf
             with a H-rich atmosphere (DA) is known to exhibit periodic brightness
             variations. Whether this cut-off in flux variations is actually due to a
             cessation of pulsation or merely due to the attenuation of any variations
             by the convection zone, rendering them invisible, is not clear. The latter
             is a theoretical possibility because with decreasing effective temperature, 
             the emergent flux variations become an ever smaller fraction of the amplitude 
             of the flux variations in the interior. In contrast to the flux variations, the
             visibility of the velocity variations associated with the pulsations is not
             thought to be similarly affected. Thus, models imply that were it still pulsating, 
             a white dwarf just below the observed red edge should show velocity variations.
             In order to test this possibility, we used time-resolved spectra of three
             DA white dwarfs that do not show photometric variability, but which have
             derived temperatures only slightly lower than the coolest ZZ Ceti variables.
             We find that none of our three targets show significant periodic velocity
             variations, and set 95\% confidence limits on amplitudes of 3.0, 5.2, and
             8.8\,km\,s$^{-1}$. Thus, for two out of our three objects, we can rule out
             velocity variations as large as 5.4\,\kms\ observed for the strongest mode
             in the cool white dwarf pulsator \object{ZZ Psc}. In order to verify our 
             procedures, we also examined similar data of a known ZZ Ceti, \object{HL Tau 76}. 
             Applying external information from the light curve, we detect significant 
             velocity variations for this object with amplitudes of up to 4\,km\,s$^{-1}$. 
             Our results suggest that substantial numbers of pulsators having large velocity 
             amplitudes do not exist below the observed photometric red edge and that the 
             latter probably reflects a real termination of pulsations.
   }
   \authorrunning{Kotak}
   \titlerunning{Pulsating or not?}
   \maketitle

\section{Introduction}
\label{sec:intro}
In spite of advances in theoretical formulations and observational capabilities, 
the details of the inner workings of the cool H-rich pulsating white dwarfs 
(DAVs or ZZ Cetis) continue to elude us. 

Early attempts to explain the nature of the driving of the pulsations
of ZZ Cetis postulated that driving occurred via the $\kappa$-mechanism 
-- i.e. in a manner akin to that in Cepheids and $\delta$-Scuti stars --
\citep[e.g.][]{danv:81,dzak:81,wing:82} 
thereby ignoring the effects of pulsation on the convection zone (also 
known as the ``frozen-in'' approximation) even though radiative flux transport
is negligible in these regions.
\citet{brick:91} realised that the response of the convection zone to the
perturbation is almost instantaneous ($\sim 1$\,s) compared to the periods of the 
g-modes (hundreds of seconds). He found that as a result, the convection
zone itself can drive the pulsations: part of the flux perturbations entering the 
convection zone from the largely radiative interior are absorbed by the convection 
zone and are released half a cycle later. This interaction between the perturbation 
and the response of the convection zone drives the g-modes. 

Within the framework of this ``convective-driving'' mechanism, recently confirmed analytically
by \citet{gw:99a}, the blue edge of the instability strip i.e. the hottest temperature for 
which pulsations are excited and at which pulsations are expected to be observable, is set by 
the condition that $\omega \tau_{c} \approx 1$ for radial order, {\em n \/} = 1 and spherical 
degree $\ell=1$; $\omega$ is the radian frequency and $\tau_{c}$ the thermal time constant of 
the convection zone. \citep[Note that $\tau_{c}$ is different from the `global' thermal 
time scale, $t_{t}$ in \citet{brick:91}. Where necessary, we follow the notation of][] 
{gw:99a}.  

The location of the red edge is less clearly defined. As long as a mode is driven, its
intrinsic amplitude is likely to remain roughly constant as it is most likely set by
parametric resonance with stable daughter modes \citep{wg:01}.
As the white dwarf cools, the depth of the convection zone increases, and damping 
in the shear layer at the base of the convection zone becomes stronger.
At some stage, damping will exceed the driving due to the convection zone. 
This sets the physical red edge, beyond which pulsations are no longer excited.

Observationally, though, the red edge may appear at {\em higher\/} temperatures. This is 
because, as the convection zone deepens, the amplitude of flux variations at the photosphere
becomes an ever smaller fraction of the amplitude in deeper layers; the convection 
zone acts as a low-pass, frequency-dependent filter leading to a reduction in flux given 
by \citep{gw:99a}
\begin{equation}
 \left(\frac{\Delta F}{F}\right)_{\rm ph} = \left(\frac{\Delta F}{F}\right)_{\rm b}\frac{1}
 {\sqrt{1 + (\omega\tau_{\rm c})^{2}}}
\label{eq:dfof}
\end{equation}
where the subscripts ``ph'' and ``b'' refer to the photospheric flux variation and 
that at the base of the convection zone respectively, for a particular eigenmode.
From Eq. (\ref{eq:dfof}) one expects that the observed amplitude of a 
given mode will decrease smoothly with increasing $\tau_{\rm c}$, and thus with
decreasing temperature. This might seem inconsistent with the observations, which 
show a rather sharply defined red edge. However, $\tau_{\rm c}$ depends very steeply 
on temperature, and hence it may also be that the observations do not yet resolve the 
transition from easily to barely visible \citep[see Fig. 8 in][]{wg:99}.

Given the uncertainties in the physical processes leading to damping, it is not
clear whether the observed red edge corresponds to the physical red edge, or
whether it is just an apparent red edge where pulsations are still driven, but
do not give rise to observable flux variations at the photosphere. The implication,
then, is that a perfectly constant white dwarf might actually still be pulsating!

Theoretical uncertainties aside, there are also uncertainties in 
interpreting the observations. Indeed, over the years, the location and 
extent of the observed ZZ Ceti instability strip have undergone several 
transformations \citep[e.g.,][]{greens:82,robetc:95,giov:98}. 
This is due in part to the difficulty in accurately determining the
atmospheric parameters of objects having convectively unstable
atmospheres, as some version of the mixing length prescription
usually has to be assumed. Additionally, at optical wavelengths, 
the Balmer lines attain their maximum strengths in the middle of 
the instability strip. The unfortunate consequence is that varying 
the atmospheric parameters gives rise to only slight changes
in the appearance of the spectra, making it difficult to uniquely
fit observed spectra \citep[e.g.,][]{kv:96}.
Supplementary constraints in the form of uv spectra, parallaxes,
gravitational redshifts, have almost become a prerequisite.

Most studies to date have only considered flux variations. However, velocity
variations are necessarily associated with these flux variations and though
small -- of the order of a few \kms\ -- have nevertheless been measured in
ZZ Ceti white dwarfs: securely in ZZ Psc \citep{vkcw:00}, and probably also
in \object{HS 0507+0434B} \citep{kotak:02}. That there are negligible vertical 
velocity gradients in the convection zone due to damping by turbulent viscosity 
is a central tenet of the convective-driving theory \citep{brick:90,gw:99b}.
Indeed for shallow convection zones, this simplification has been 
shown to hold in the 2D hydrodynamical simulations of \citet{gautschy:96}.
This means that the horizontal velocity is nearly {\em independent}
of depth in the convection zone so that although photometric variations 
become difficult to detect around the red edge of the instability strip 
as mentioned above, velocity variations pass virtually undiminished
through the convection zone. 

\begin{table*}[htb]
\caption[]{Observations}
\fontsize{8.5}{10}\selectfont
\label{tab:obs}
\begin{centering}
\begin{tabular}{lcccccccccc}
\hline
             & WD  &  \Teff  & $\log g$ & V & Start & End & Exposure & No. of & Scatter & Measurement \\
             & number &  (K)    &          &   & U.T.  & U.T. & Time (s) & frames & (\kms)  & error (\kms)  \\
\hline
 HL Tau 76 (s)& 0416+272 &   11440  & 7.89 $\pm$ 0.05 & 15.2 & 12:58:55 & 14:59:06 & 20  &   213    & 9.0  & 6.7  \\
 G 1-7 (s)    & 0033+013 &   11214  & 8.70 $\pm$ 0.06 & 15.5 & 08:37:49 & 10:39:50 & 74  &  \phn83  & 9.8  & 7.7 \\
 G 67-23 (s)  & 2246+223 &   10770  & 8.78 $\pm$ 0.03 & 14.4 & 07:33:57 & 08:48:15 & 74  &  \phn50  & 3.5  & 2.4  \\
 G 126-18     & 2136+229 &   10550  & 8.17 $\pm$ 0.05 & 15.3 & 04:46:06 & 06:27:25 & 30  &   136    & 1.4  & 7.6 \\
 G 126-18 (s) &          &          &                 &      & 06:58:18 & 08:58:03 & 74  &  \phn80  & 5.2  & 4.0  \\
\hline
\end{tabular}
\tablecomments{The effective temperature and $\log g$ for each of the 3 red edge
objects are taken from \citet{kep:95} and are inferred from optical data only
using the ML2, $\alpha$=1 prescription for describing the convective flux.
For HL Tau 76, these quantities are taken from \citet{berg:95} and are
constrained using both uv and optical spectra for ML2/$\alpha$=0.6.
The scatter due to wander is calculated at a representative wavelength of
4341\,{\AA} i.e. at H$\gamma$. The measurement error is the typical internal
error on the line-of-sight velocity associated with the fits to the line profiles as
described in Sect. \ref{sec:vels}. ``(s)'' refers to the Oct. 1997 service
run data. The number of frames refers to the number of {\em useful\/} frames.}
\end{centering}
\end{table*}

Observationally, the consequence would be that the putative, perfectly 
constant white dwarf might reveal its pulsating nature by velocity variations.
If objects below the red edge follow the same trend as the known pulsators 
i.e. longer pulsation periods and higher amplitudes with decreasing effective 
temperature \citep{clem:93}, then we expect the highest velocity amplitudes at the 
longer ($\ga$ 600\,s) periods. For ZZ Psc (a.k.a. G 29-38), a well-known pulsator
close to the red edge, \cite{vkcw:00} measured a velocity amplitude of 
5.4\,\kms\ for the strongest mode at 614\,s. For all our objects, which are cooler, 
we expect velocity amplitudes at least as large as those measured for ZZ Psc -- lower 
values would constitute a non-detection.

Armed with this testable theoretical expectation, we look for variations in the 
line-of-sight velocities in objects that lie just below the photometric red edge of the 
instability strip. Our three targets are chosen -- subject to visibility constraints 
during the scheduled run -- from the list of \citet{kep:95} who find a handful
of non-pulsating ZZ Cetis close to the red edge of the instability strip.
\citet{kep:95} inferred relatively high masses for two of our three targets
(see Table \ref{tab:obs}). This inference, though dependent on the assumed
convective prescription, would imply that these objects have instability strips at higher
temperatures.

In addition to the above, we search for line-of-sight velocity variations
in the first ZZ Ceti to be discovered, HL Tau 76 \citep{land:68}, using exactly the
same instrumental setup and reduction techniques so that it serves as a
comparison case. Treating it subsequently as a ZZ Ceti variable (Appendix \ref{sec:hlt})
allows us to indirectly constrain the spherical degree of the eigenmodes.

\section{Observations and Data Reduction}

Time-resolved spectra of three alleged non-pulsators, \object{G 1-7}, 
\object{G 67-23}, and \object{G 126-18} \citep{kep:95} and one known 
ZZ Ceti-type white dwarf, HL Tau 76, were acquired using the Low Resolution 
Imaging Spectrometer \citep{oke:95} on the Keck II telescope
during a service run on the nights of $18^{\mathrm{th}}$ and
20$^{\mathrm{th}}$ October 1997. In addition to the above, we took
time-resolved spectra of G 126-18 during an observing run with the same 
instrument and set-up on 11$^{\mathrm{th}}$ December 1997. Details of the 
observations together with other quantities are summarised in Table 
\ref{tab:obs}. An 8$\farcs$7 slit was used with a 600 line mm$^{-1}$ 
grating, the resulting dispersion being $\sim$1.25\,{\AA}\,pix$^{-1}$. For 
the service run, the seeing was around 1$\farcs$0, yielding a resolution
of 6\,{\AA}, while the December 1997 data were taken in slightly worse 
seeing conditions (1$\farcs$2), corresponding to a resolution of about 
7\,{\AA}. A problem with the Active Control System meant that 22 frames 
of G 67-23 taken during the beginning of the run had to be discarded and
4 frames taken at the very end of the run also had to be disregarded as
the position of the target in the slit seemed to have altered. The last
frame of HL Tau 76 was not used as it was clear that light was being lost. 
For the observations taken during the service run, we were unfortunately 
unable to obtain a sufficient number of halogen frames so we decided to forgo 
flatfielding the spectra altogether.

Shifts in the positions of the Balmer lines are introduced by instrumental 
flexure and differential atmospheric refraction. As in \citet{vkcw:00}, we 
corrected for the former using the position of the \ion{O}{i} $\lambda$\,5577\,{\AA} 
sky emission line, and for the latter by computing the wavelength-dependent shifts 
using the recipe of \citet{stone:96}. As the spectra were taken through a wide slit, 
the positions of the Balmer lines will also depend on the exact position of the target 
in the slit. In principle, this position should be tied to the position of the guide 
star, but in practice we found that there was substantial random jitter, probably due 
to guiding errors and differential flexure between the guider and the slit.

One way of accounting for the random movements in the slit is to in effect
`tag' the movements of the target to those of a reference star that can also
be accommodated in the slit. For the Dec. 1997 observations, the slit was set 
at a position angle such that a G-type star was in the slit together with G 126-18. 
The wandering of G 126-18 could then be corrected for using the positions of H$\beta$ 
of the former as fiducial points. 

An estimate of the scatter due to random wander in the slit can be obtained by 
measuring the shifts along the slit of the spatial profiles and translating these 
into scatter in the dispersion direction. To do this, we simply fit Gaussian
functions to the spatial profiles, and fit a low order polynomial to the resulting 
centroid positions (if no reference star was available). The standard deviation of 
the centroid positions from the smooth curve is then a proxy of the scatter due to 
wander in the slit. We compute this scatter at the representative wavelength of 
H$\gamma$. The resulting estimates are shown in Table \ref{tab:obs}. As expected, 
the use of a reference object for G 126-18 dramatically lowers the scatter in the 
measured line-of-sight velocities (see Table \ref{tab:obs}). However, the value 
for G 67-23 shows that commensurate results can also be obtained without a local 
calibrator if the observing conditions and guiding are sufficiently stable.

\begin{figure*}
\begin{center}
 \mbox{
\epsfysize=0.34\textwidth \epsfbox{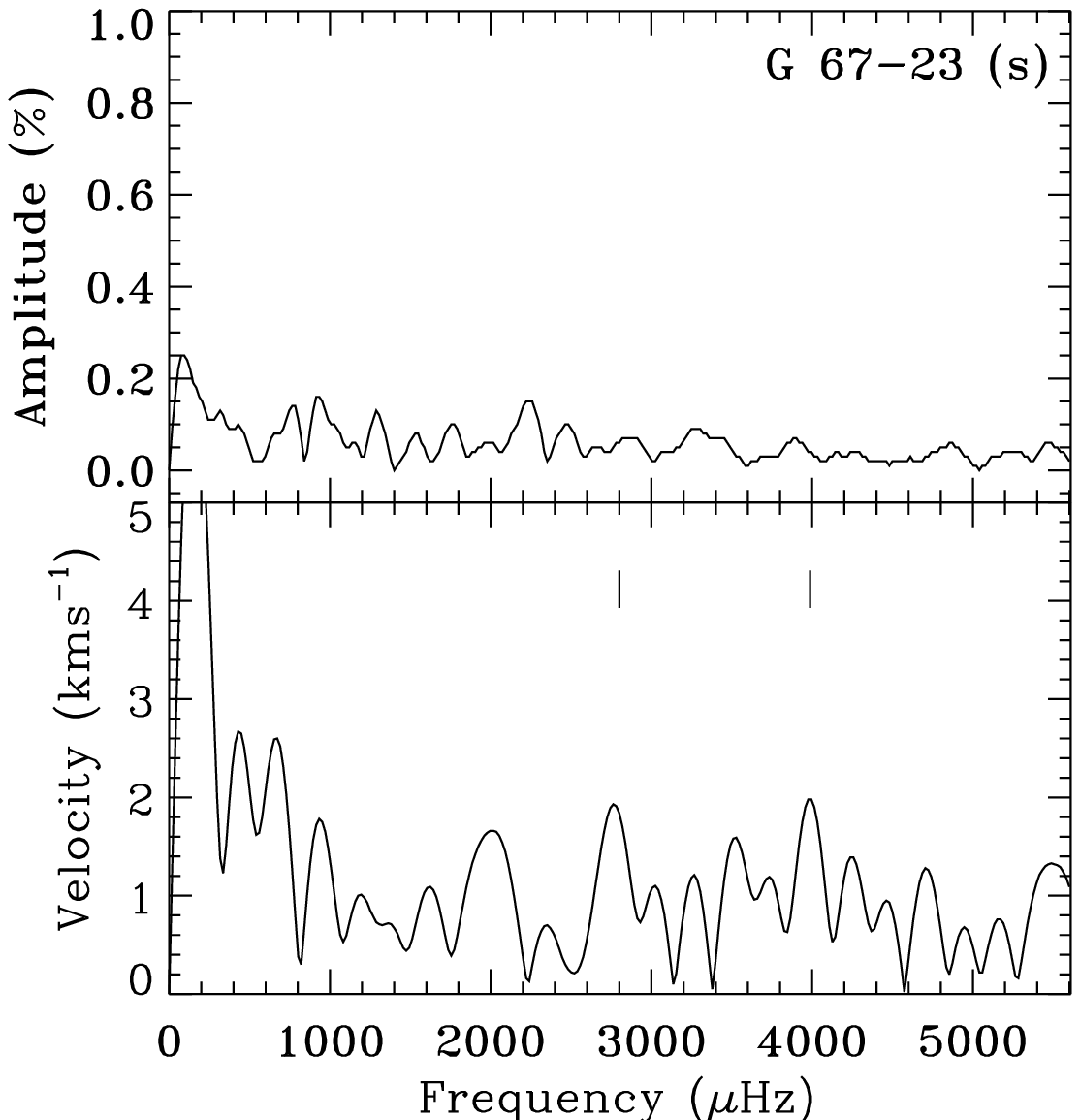}
\epsfysize=0.34\textwidth \epsfbox{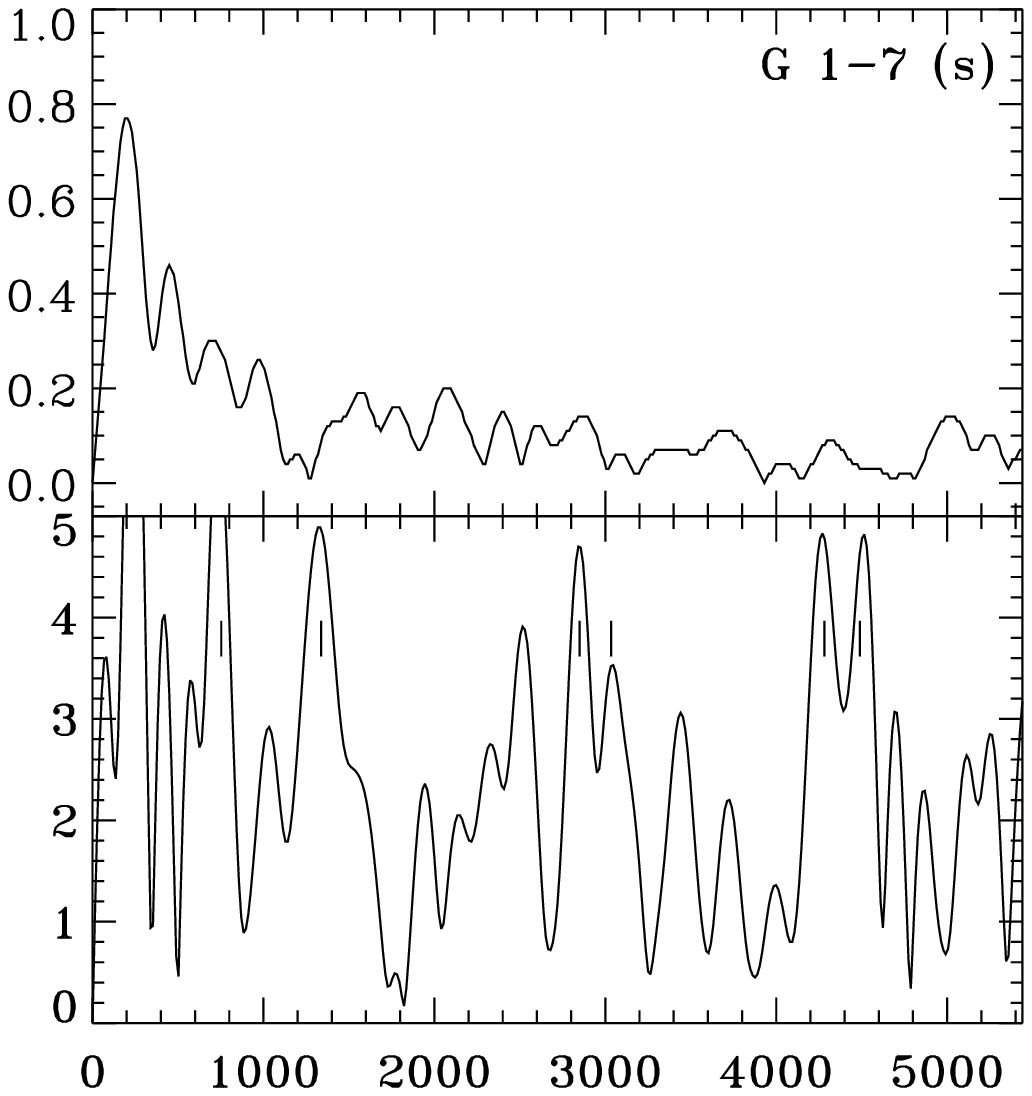}
\epsfysize=0.34\textwidth \epsfbox{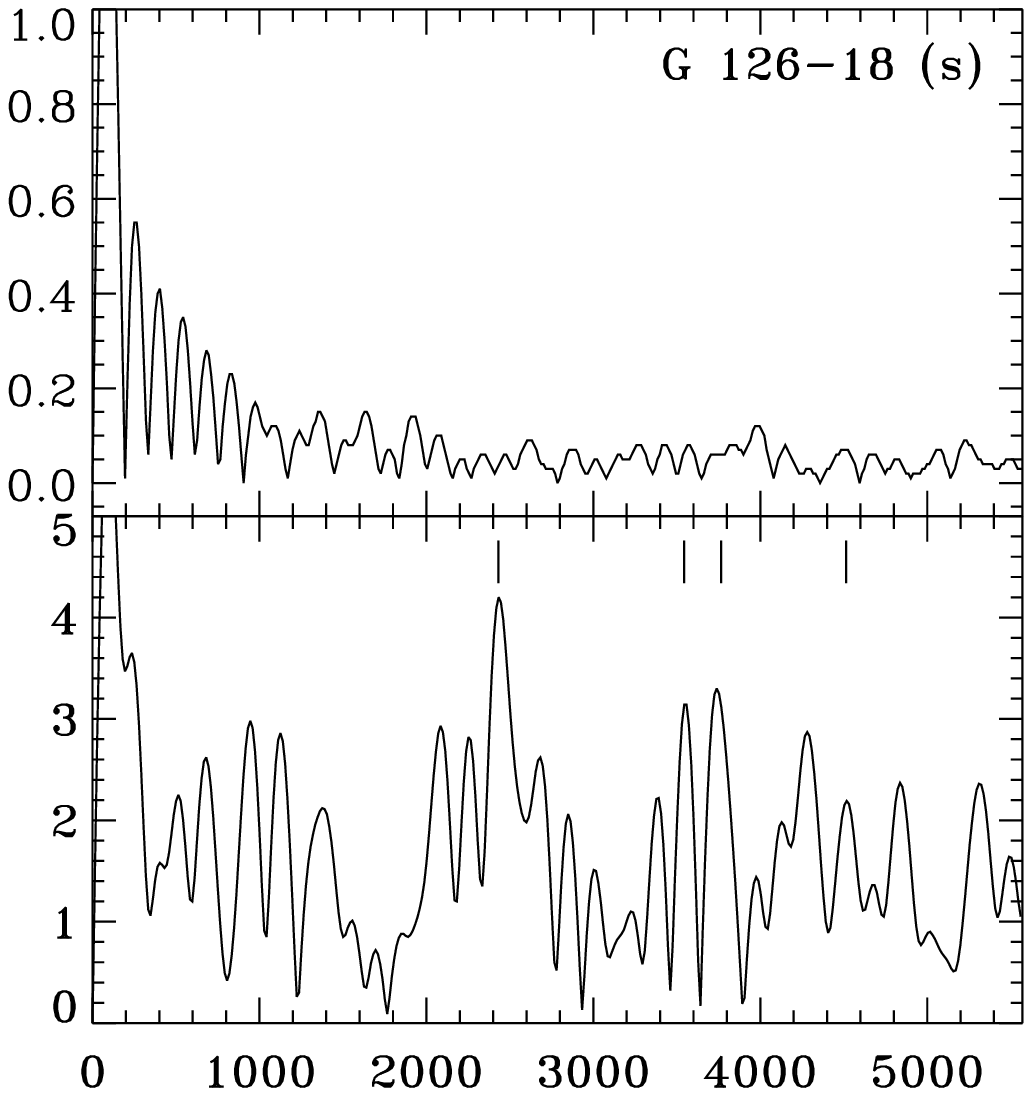}
  }
\mbox{
\epsfysize=0.41\textwidth \epsfbox{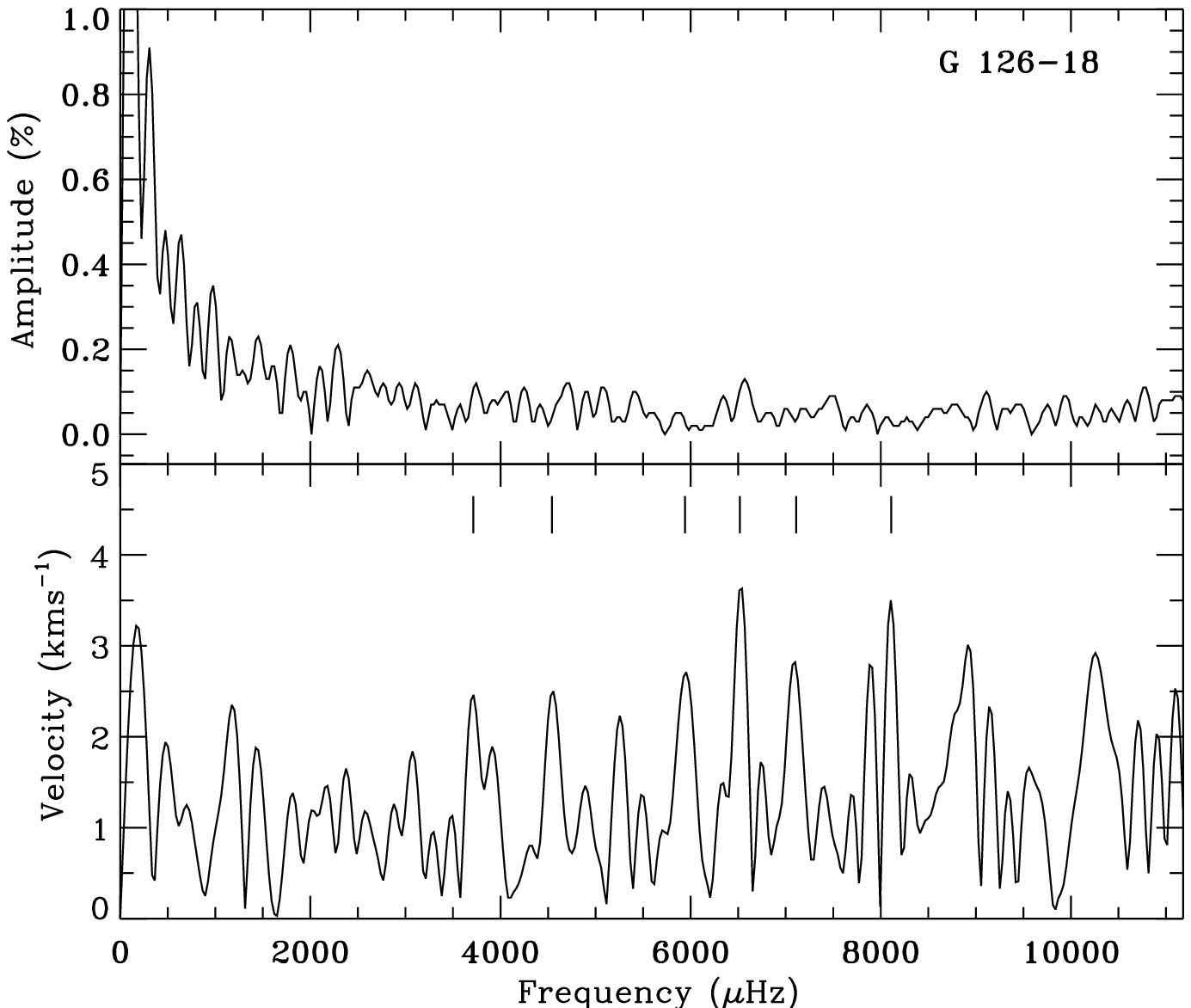}
\epsfysize=0.41\textwidth \epsfbox{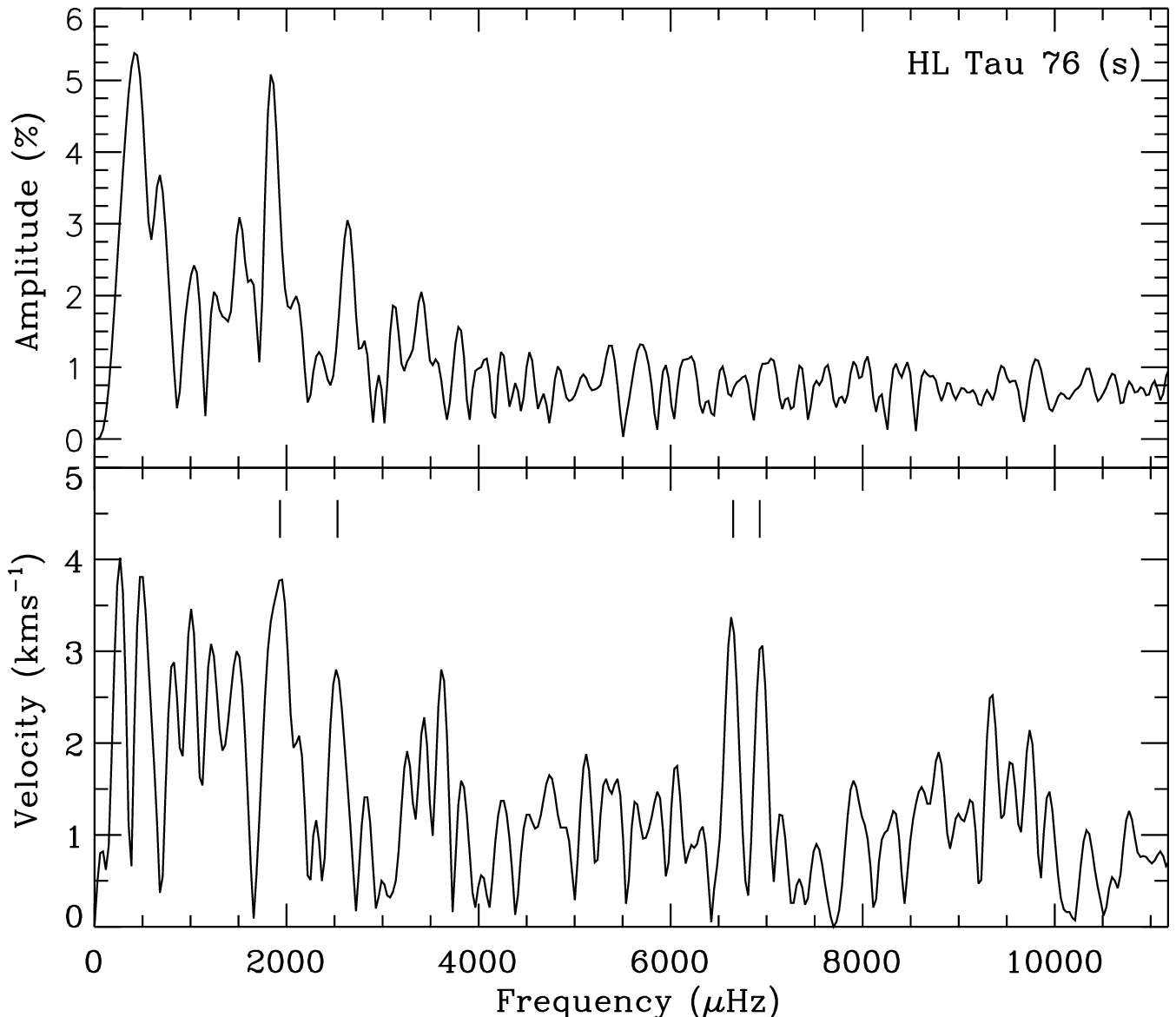}
}
\end{center}
\caption{Fourier transforms of the light (top panels) and velocity curves (bottom
panels) for all objects. The marked peaks have been included in the Monte Carlo simulations
in Sect. \ref{sec:sign}. We find no significant peaks in either the flux or the velocity FTs
for G 1-7, G 126-18, and G 67-23. All FTs are shown up to the Nyquist Frequency except
that of HL Tau 76 for clarity (see also caption to Fig. \ref{fig:hlflxovelzz}).}
\label{fig:bigfive}
\end{figure*}

\section{Lightcurves}
\label{sec:lc}
Lightcurves for all objects were constructed by dividing the line-free
region of the continuum between $\sim$\,5300 -- 5700\,{\AA} by the average 
of that region. The Fourier transforms of the light curves were calculated 
up to the Nyquist frequency and are shown in Fig. \ref{fig:bigfive}.
No clear periodicities are evident for G 67-23, G 1-7, and G 126-18, as 
expected based on the limits of $0.25\%$, $0.37\%$, and $0.46\%$ respectively
set by \cite{kep:95}.

\section{Line of sight velocities}
\label{sec:vels}

Velocity curves were constructed by measuring the
Doppler shifts of three Balmer lines: H$\beta$, H$\gamma$, and H$\delta$.
This was done by fitting a sum of a Gaussian and Lorentzian profile and
approximating the continuum by a line. The centroids of the Gaussian and
Lorentzian functions were forced to be the same. The wavelength intervals
chosen for our fits were $\sim$\,388\,{\AA}, 300\,{\AA}, and 164\,{\AA} centred 
approximately on H$\beta$, H$\gamma$, and H$\delta$ respectively. We
computed the velocity curve by simply averaging the velocities derived from
each line.

The measurement errors are of the order of the scatter due to random wander
in all but one of the series of spectra (Table \ref{tab:obs}). The one exception 
is the series taken of G~126-18 in Dec. 1997, where the scatter is much
smaller because of the tie to a reference object. 
Our measurements are most precise for G 67-23. We include our estimates for the
scatter due to random wander in the least squares fit of the velocity curve
(see below); the resulting fits all have $\chi_{\rm red}^2 \simeq 1$.

The Fourier Transforms of the velocity curves are shown in Fig. \ref{fig:bigfive}.
None of the red edge objects show a clear peak well above the noise level at
any frequency. 

For HL Tau 76, two of the stronger peaks have corresponding peaks in the light curve. 
However, there are also two other peaks evident in Fig. \ref{fig:bigfive} at 6651\,$\mu$Hz 
and 6928\,$\mu$Hz that are not present in the light curve at amplitudes greater 
than $0.9\pm0.8$\% and $0.6\pm0.8$\% respectively i.e. well below our detection limit.
We will return to this point in Appendix \ref{sec:hlt}.

Although it is clear from Fig. \ref{fig:bigfive} that none of our objects show
a large ($\simgt 5$\,\kms) peak at any frequency, given the disparities in the
quality of our various data sets we must carry out tests in order to be able to place
meaningful limits. We describe below two different Monte Carlo tests we devised to 
obtain quantitative estimates.

\subsection{Significance of possible detections}
\label{sec:sign}

We first looked for periodicities in each of the velocity curves 
by concentrating on the few peaks of relatively high amplitude in the 
Fourier transform of the velocity curves and fitting these successively with 
sinusoids, the frequencies, amplitudes, and phases all being free parameters. 
We included a low order polynomial in the fit to remove slow variations for 
all objects except G 126-18 (Dec. 1997) for which the calibration with respect 
to the reference star will have already removed any slow variations.

The first test we carry out aims to determine the likelihood of the peaks seen
in the Fourier Transforms (Fig. \ref{fig:bigfive}) as being simply due to random 
noise. We shuffle the measured velocities randomly with respect to their measurement
times after removing slow variations by fitting a 3$^{\mathrm{rd}}$-order
polynomial for all objects (except G 126-18, Dec. 1997). We then determine 
the amplitude of the highest peak in the Fourier transform at any frequency 
between 700\,$\mu$Hz up to the Nyquist frequency. (This choice reflects 
an extended ZZ Ceti period range). We repeat this procedure 1000 times, counting 
the number of artificial data sets in which the highest peak exceeded our maximum 
measured $A_V$. The higher this number, the higher the chance that even the 
strongest peak is not genuine. Table \ref{tab:mcvel} shows our results. 
The strongest peaks in all the red edge objects can be attributed to noise.
Only in HL Tau 76, the one known pulsator, is there an indication that the
velocity signal is real. We show in Appendix \ref{sec:hlt} that external 
information (from the light curve) is necessary to confirm this. 

\begin{table}[!hb]
\caption[]{Results from Monte Carlo simulations of velocity amplitudes}
\fontsize{7.6}{9}\selectfont
\label{tab:mcvel}
\begin{centering}
\begin{tabular}{lcccccc}
\hline
\multicolumn{3}{r}{Largest Peak in F.T.} &
\multicolumn{1}{c}{} &
\multicolumn{1}{c}{Detection} \\

 Object   & $\nu$ & $A_{V}$  & P($>A_{V})$ &  Limit ($\sim$95\%) \\
          & ($\mu$Hz)    & (\kms)          & (\%)    &  (\kms) \\
\hline
    G 1-7 (s)        & \phn753  & 6.0 $\pm$ 1.5  & 42   &   8.8 \\
    G 126-18         &  6518    & 3.4 $\pm$ 1.0  & 67   &   5.2 \\
    G 126-18 (s)     &  2430    & 4.1 $\pm$ 1.2  & 35   &   5.9 \\
    G 67-23 (s)      &  3988    & 2.0 $\pm$ 0.7  & 60   &   3.0 \\
    HL Tau 76 (s)    &  1933    & 4.1 $\pm$ 1.0  & 17   &   5.1 \\
\hline
\end{tabular}
\end{centering}
\tablecomments{$A_V$ is the maximum measured velocity amplitude for each object
at the frequency ($\nu$) specified in the first column. P($>A_{V}$) is the 
percentage probability that the largest peak (i.e. of amplitude $A_{V}$) is present 
solely by chance. The third column lists the velocity amplitude necessary to 
constitute a detection at the 95\% confidence level -- at any frequency --
as estimated from our second Monte Carlo test (see also Fig. \ref{fig:maxfakepeak}). 
Note that the frequency corresponding to the maximum velocity peak is different for 
the two G 126-18 data sets.} 
\end{table}

\begin{figure}
\plotone{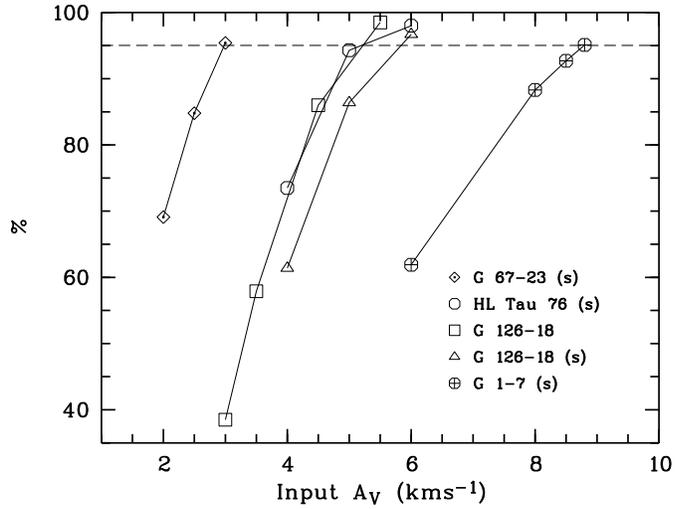}
\caption{Probabilities that the velocity signals with the given amplitudes
that would have been detected in the different data sets had they been present.
The dashed line shows the 95\% confidence level, at or above which measured 
velocities would be taken to be significant. For our best data set (G 67-23),
we can exclude periodic signals with amplitudes greater than 3\,\kms\ with 95\%
confidence.} 
\label{fig:maxfakepeak} 
\end{figure}

Our second objective is to determine the minimum velocity amplitude in each data set
for which we could have had a high ($\ga$ 95\%) degree of confidence that, had
such a signal been present, we would have detected it.
To do this, we inject an artificial signal in our data of a given amplitude
and random frequency and phase. The random frequencies are chosen from
700\,$\mu$Hz up to the Nyquist frequency, while the phases are chosen from 0 to $2\pi$.
We repeat the procedure 1000 times and count the number of artificial data sets in
which the highest peak in the Fourier transform is located within a range corresponding
to the input random frequency plus or minus the frequency resolution of the data.
We repeat the procedure, varying the input amplitude, until we recover the 
artificial peak as the maximum peak in at least 95\% of the trials within the
expected frequency range. We expect that the velocity amplitude we obtain in this
manner will be somewhat higher than the detection limit derived above. This is 
because for peaks having an amplitude exactly at the detection limit, there is an 
equal chance of a noise peak either enhancing or diminishing it. Thus for a 
peak exactly at the detection limit, we cannot expect $>$50\% confidence. This test 
implicitly assumes that none of the signal in our Fourier transform is real.
For HL Tau 76, we show in the appendix that the higher peaks in the velocity
Fourier Transform that are coincident with peaks in the Fourier Transform of the
light curve are real. Hence, the limits for any signal not associated with flux
variations will be slightly lower than those inferred from Fig. \ref{fig:maxfakepeak}.

The results of the above simulation are displayed in Fig. \ref{fig:maxfakepeak}.
We can rule out velocity amplitudes larger than 5.4\,\kms\ -- as detected by 
\citet{vkcw:00} for ZZ Psc -- for all objects except G 1-7. Our best limit for a 
minimum significant amplitude (3\,\kms) comes from G 67-23.

\section{Discussion and Conclusions}

We have analysed velocity curves of three white dwarfs below the photometric
red edge of the instability strip to check whether the observed red edge corresponds
to the physical one. For ZZ Psc, a white dwarf close to the red edge but still within 
the instability strip, \citet{vkcw:00} found a maximum velocity amplitude of 
$5.4 \pm 0.8\,$\kms. We had expected larger modulations but for two objects we can 
exclude the presence of signals as strong as those in ZZ Psc with high confidence; for 
one of these two, G 67-23, we can exclude the presence of any signal greater than 3\,\kms.

The above indicates that pulsations actually cease below the observed
red edge. How secure is this conclusion? We see three weaknesses.
The first is that in principle, destructive beating between different 
components in a multiplet might conceal a real signal in our short data sets. This, 
however, would be unlikely if multiple real signals were present. Furthermore, two data 
sets for the same object (G\,126-18) taken during different observing runs are unlikely 
to be adversely affected. Two of the modulations found in G 126-18 (viz. 220\,s 
and 269\,s) seemed to be consistent with being at the same frequency and of the same 
amplitude -- within the errors -- for the two different observing runs. We carried 
out further Monte Carlo simulations to determine the likelihood of finding relatively 
strong peaks at the same frequency in two separate data sets. On the basis of these 
simulations, we conclude that these peak coincidences are almost certainly due to 
chance: even for the best case, the peaks at 265\,s and 269\,s, there is a 7\% 
probability of a chance coincidence.

The second weakness is that our sensitivity turned out to be more 
marginal than we had hoped. We included one known ZZ Ceti, HL Tau 76, in the 
sample.  Treating it as a red-edge object afforded a check on our methods.
However, even for this bona-fide pulsator, we found that the velocity
curve on its own was not sufficient to demonstrate velocity variations. A hint 
of a real signal was present, but this could only be confirmed by imposing 
external information from the light curve, information we lacked for the red-edge 
objects. In the process, HL Tau 76 was added to the select group of ZZ Cetis for
which velocity variations have been detected.

The above might suggest that our experiment was simply not 
sensitive enough. One has to keep in mind however, that for the 
different objects rather different sensitivities were reached. 
For one object, G 67-23, we can exclude the presence of velocity
amplitudes as small as 3\,\kms\ i.e. even smaller than the 4\,\kms\ peak
corresponding to strongest mode in HL Tau 76.
Furthermore, on theoretical grounds alone, a velocity amplitude somewhat
smaller than that of ZZ Psc is expected for HL Tau 76 given that it has a
slightly higher temperature and its strongest mode, a slightly shorter period.
This expectation is supported by the fact that the velocity amplitude
of the dominant mode is in turn larger than that found for HS~0507+0434B,
which is somewhat hotter still and has an even shorter dominant periodicity.
On the same grounds, one would expect, as mentioned earlier, the red-edge
objects to have larger velocity amplitudes than those observed in ZZ~Psc,
which, if present, would have been seen in two of the three objects.\footnote{We 
remind the reader that the amplitude which a driven mode reaches is not expected 
to depend on temperature, but on resonance conditions with daughter modes; see
\citet{wg:01}.}

The third weakness is perhaps the most severe: 
G\,1-7 and G\,67-23 have relatively high inferred masses, which implies 
that these objects would have ceased to be pulsationally active at a higher
effective temperature than that expected for a typical $0.6\,M_\odot$ 
white dwarf, and are therefore not that close anymore to the red edge. 
At the same time, G\,126-18, which has a normal mass, is the coolest of our 
three observed targets. Note though that the derived effective temperatures 
and surface gravities depend on the treatment of convection.

In summary, while not ironclad, our results indicate that pulsations
have actually ceased below the observed red edge rather than having
become photometrically undectable, and that the theoretical
expectation of an extended instability strip, beyond the observed red
edge is flawed. In order to settle the issue observationally, more sensitive 
measurements on objects closer to the red edge would be worthwhile; theoretically, 
detailed hydrodynamic modelling might go some way towards understanding the
interplay of the various physical processes in defining the red edge.

\begin{acknowledgements}
We thank the referee, Detlev Koester, for his comments,
and are grateful to Y. Wu for answering our many questions.
R.K. would also like to thank J.S. Vink for encouragement and 
H-G. Ludwig for useful discussions. 
We also acknowledge support for a fellowship of the Royal Netherlands
Academy of Arts and Sciences (MHvK) and partial support from the
Knut och Alice Wallenbergs Stiftelse (RK). This research has made use 
of the SIMBAD database, operated at CDS, Strasbourg, France.
\end{acknowledgements}

\appendix
\section{HL Tau 76}
\label{sec:hlt}

\begin{figure}[!t]
\plotone{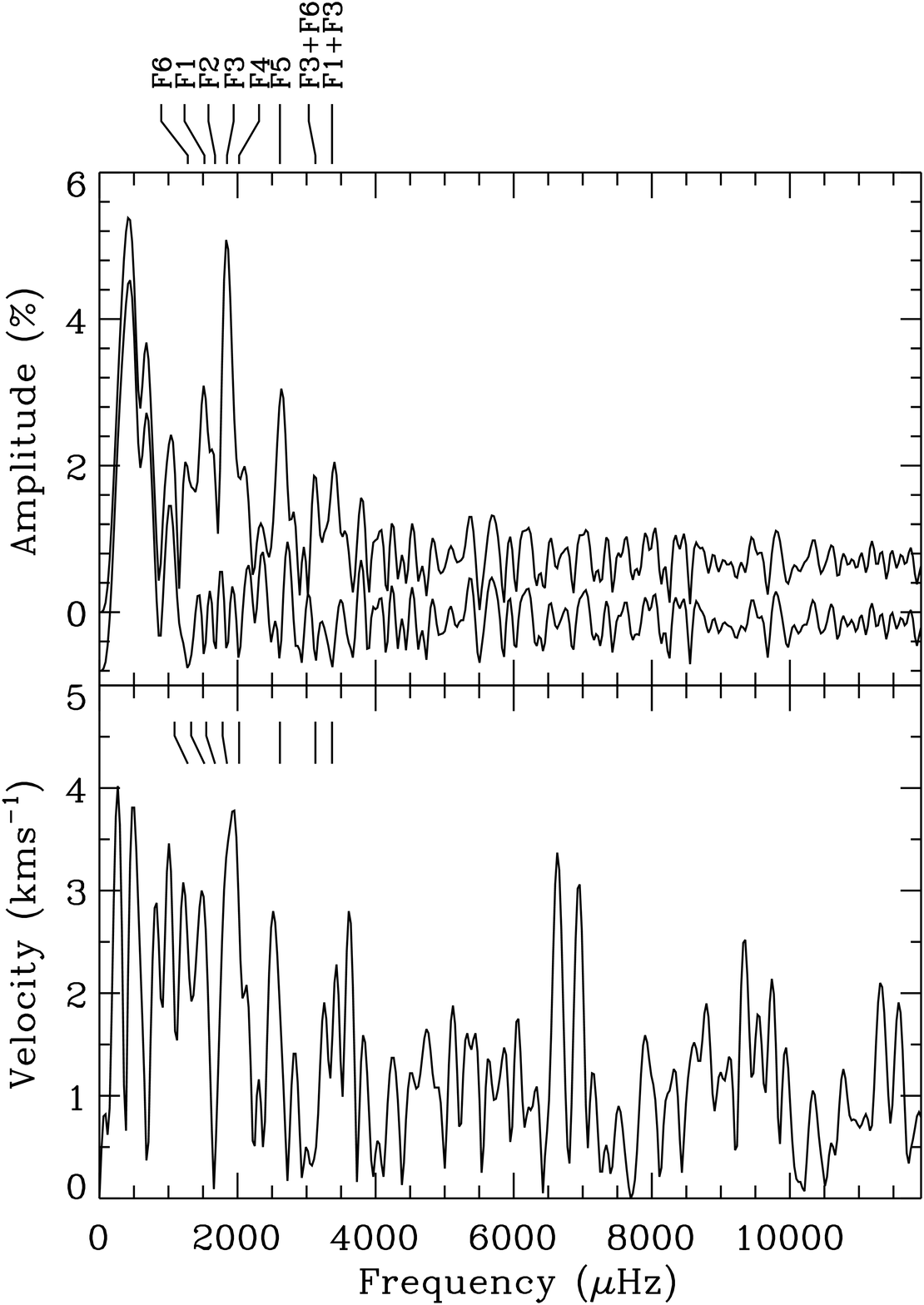}
\caption{Fourier transform of the light curve (a) and (b) of the line-of-sight
velocity variations shown up to 11900\,$\mu$Hz for HL Tau 76.
There are no peaks greater than 1\% and 2.5\,\kms\ longward of
11900\,$\mu$Hz for (a) and (b) respectively.
The lower curve of the upper panel shows the residuals (offset by $-$0.8\%)
after fitting sinusoids at the frequencies indicated.}
\label{fig:hlflxovelzz}
\end{figure}
\begin{figure}[!t]
\plotone{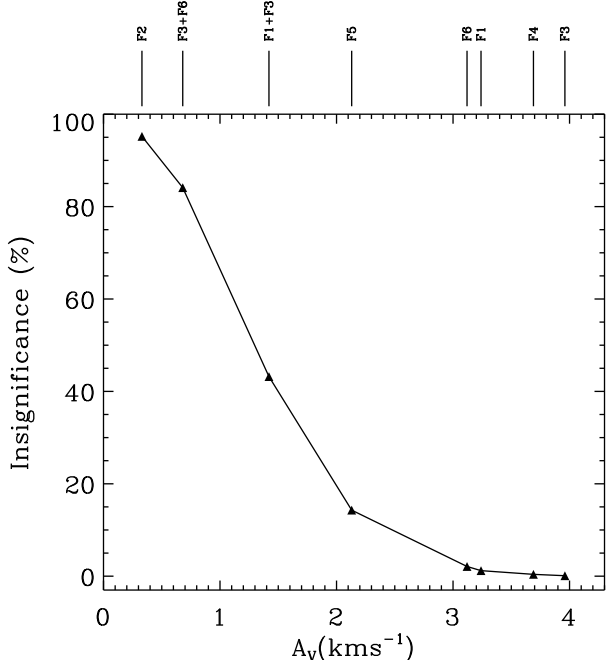}
\caption{Results of the Monte Carlo simulation described in Appendix \ref{sec:hlt} i.e.
fitting randomly shuffled velocity curves of HL Tau 76 with frequencies fixed to those
obtained from the WET data. The triangles mark the velocity amplitudes thus derived.
These amplitudes and their corresponding phases are listed in Table \ref{hltab}.
The figure shows that the signals at F3, F4, F1, and F6 are unlikely to be due to
chance.}
\end{figure}
The Fourier Transform of the light curve of HL Tau 76 is characteristic of that of 
the ZZ Cetis (Fig. \ref{fig:hlflxovelzz}). We fit the dominant peaks in the Fourier
transform successively with a function of the form $A\cos(2\pi ft'-\phi) + C$
where $A$ is the amplitude, $f$ the frequency, $\phi$ the phase, and $C$, a constant 
offset.
In order to reduce the covariance between the amplitudes and phases, the time axis is
computed relative to the the middle of the time series ($t' = t\,-\,13:59:01.56$ U.T.).
We fixed the frequencies to those listed by \citet{dank:96} in their preliminary
analysis of WET \citep[Whole Earth Telescope;][]{nath:90} data as these
should be more reliable than the frequencies inferred from our data set, which only
covers a short time span. We include all potential real modes listed in Table 4
of \citet{dank:96} with an amplitude $\ge 0.9\%$ except for modes at frequencies
below $\sim 1000\,\mu$Hz as this corresponds to the noisiest region of our Fourier 
transform.  The relevant quantities are listed in Table \ref{hltab}. We find that our 
measured amplitudes ($A_L$) are on average 1\% higher than those from the WET analysis. 

Comparing the flux and velocity Fourier Transforms, we note first of all, that the 
highest peak in Fourier Transform of the velocity curve coincides with the dominant 
peak in the Fourier transform of the light curve. Other such coincident peaks also 
seem to be present, albeit at a lower amplitude. Since the two peaks at 6651\,$\mu$Hz
and 6928\,$\mu$Hz in Fig. \ref{fig:hlflxovelzz}b are not present in the light curve
(at amplitudes greater than $0.9\pm0.8$\% and $0.6\pm0.8$\% respectively), we surmise 
that they are due to noise.

\begin{table*}[t]
\caption[]{Pulsation frequencies and other derived quantities from the light
and velocity curves for HL Tau 76. }
\fontsize{7.6}{10}\selectfont
 \begin{centering}
 \begin{tabular}{lclcccccc}
 \label{hltab}
   &          &           &              &            &         &       &         &          \\
\hline
Mode &   Period &  Frequency      & $A_{L}$   &   $\Phi_{L}$ &  $A_{V}$    &    $\Phi_{V}$ & $R_{V}$       & $\Delta\Phi_{V}$ \\
     &    (s)   & ($\mu$\,Hz)      & (\%)    &  (\arcdeg)     &   (\kms)    & (\arcdeg)    & (Mm\,rad$^{-1})$ & $(\arcdeg)$ \\
\hline
\sidehead{Real Modes:}
F1 &     657 &    1521  &    3.3 $\pm$  0.8 &  $-$57   $\pm$ 14 & 3.2 $\pm$     1.0 &     \phn34 $\pm$    18 &   10    $\pm$ 4   &  \phs\phn91   $\pm$ 23 \\
F2 &     597 &    1675  &    1.7 $\pm$  0.8 &  $-$91   $\pm$ 28 & 0.3 $\pm$     1.1 &    \nodata             &   \phn2 $\pm$ 6   &      \nodata    \\
F3 &     541 &    1848  &    5.6 $\pm$  0.8 &   \phs42 $\pm$  \phn9 & 4.0 $\pm$ 1.1 &     \phn84 $\pm$    15 &   \phn6 $\pm$ 2   &  \phs\phn41   $\pm$ 18 \\
F4 &     494 &    2023  &    2.6 $\pm$  0.8 &   \phs70 $\pm$ 19 & 3.7 $\pm$     1.0 &     \phn47 $\pm$    16 &      11 $\pm$ 5   &  \phn$-$24 $\pm$ 25 \\
F5 &     383 &    2614  &    3.3 $\pm$  0.8 &   \phs25 $\pm$ 14 & 2.1 $\pm$     1.0 &    144     $\pm$    27 &   \phn4 $\pm$ 2   &   \phs118  $\pm$ 31 \\
F6 &     781 &    1280  &    2.2 $\pm$  0.8 &  $-$81   $\pm$ 22 & 3.1 $\pm$     1.0 &     \phn48 $\pm$    19 &      18 $\pm$ 9   &    \phs130 $\pm$ 29 \\
\sidehead{Combinations:}
Mode   &   Period &  Frequency      & $A_{L}$   &   $\Phi_{L}$ &  $A_{V}$       & $\Phi_{V}$        & $R_{C}$      & $\Delta\Phi_{C}$ \\
F1+F3  &          297 &      3369.0 & 2.3 $\pm$ 0.8 & $-$20 $\pm$ 20 & 1.4 $\pm$ 1.0 & \nodata &  6 $\pm$ 3  & \phn$-$5 $\pm$ 26  \\
F3+F6  &          320 &      3128.0 & 2.0 $\pm$ 0.8 & $-$63 $\pm$ 24 & 0.7 $\pm$ 1.0 & \nodata &  8 $\pm$ 5  & $-$24    $\pm$ 33  \\
\hline
\end{tabular}
\tablecomments{Quantities derived from the light and velocity curves with frequencies fixed to
those listed in \citet{dank:96} whose nomenclature we follow. $A_{L,V}$ and $\Phi_{L,V}$ are the
amplitudes and phases obtained from fits to the light and velocity curves respectively.
$R_{V} = A_{V}/(2\pi fA_{L})$; $\Delta\Phi_{V} = \Phi_{V} - \Phi_{L}.$
$R_{C} = A_{L}^{i\pm j}/(n_{ij}A_{L}^{i}A_{L}^{j})$ here $n_{ij} = 2.$
$\Delta\Phi_C = \Phi_L^{i\pm j} - (\Phi_L^{i} \pm \Phi_{L}^{j})$ gives the relative phase
of the combination mode with that of its constituent real modes.}
\end{centering}
   \end{table*}

Applying the external information available from the light curve to the velocity
curve, we infer the line-of-sight velocities in HL Tau 76 by fixing the frequencies to
those obtained from the light curve. The resulting values are listed in Table \ref{hltab}.

In order to subject these velocity amplitudes to similar tests as described
in Sect. \ref{sec:sign}, we carry out another Monte Carlo simulation, this time
taking into account the external information available from the light curve.
We fit the shuffled data sets with frequencies {\em fixed\/} to those observed in 
the light curve. Repeating this process 1000 times, we find that the peak in the
velocity FT at F3 -- the strongest mode in the light curve -- now has a 0.1\% 
chance of being a random peak. Clearly, this improvement c.f. Table \ref{tab:mcvel} 
is due to the imposition of additional information.

Following the notation of \citet{vkcw:00}, we derive the relative velocity to
flux amplitude ratios and phases, $R_V$ and $\Delta\Phi_V$.
The strongest mode, F3, has a period of 541\,s that is intermediate between
that of HS 0507+0434 \citep[355\,s,][]{jord:98} and ZZ Psc \citep[614\,s,][]{vkcw:00}.
The measured velocity amplitude at F3 also lies between that of HS 0507+0434 
\citep[2.6$\pm$1.0\,\kms,][]{kotak:02} and ZZ Psc \citep[5.4$\pm$0.8\,\kms,][]{vkcw:00}.
The values of $R_V$ for all of the modes are consistent with each other and with having
a spherical degree ($\ell$) of 1. However, we note with interest that the one mode in ZZ Psc
identified as having $\ell=2$ by \citet{cvkw:00}, was at a period of 776\,s which is very close 
to the period of our F6. Even though the differences in $R_V$ are not formally significant,
that F6 has the largest $R_V$ is consistent with it having a higher $\ell$ value, as
flux variations for higher $\ell$ values suffer stronger cancellation while the line-of-sight
velocities do not \citep[e.g.][]{vkcw:00}. Comparison with the pulsation periods of ZZ Psc
is justified; indeed, \cite{dank:96} comment on the striking similarities in the modes
observed for these two white dwarfs.
A more definitive $\ell$-identifcation might be possible from analysing the wavelength
dependence of pulsation amplitudes or from a much longer time series which is able to
resolve multiplets within a period group.

\end{document}